\documentclass [amssymb,amsmath, twocolumn, showpacs]{revtex4-1} % twocolumn, 
\usepackage{graphicx,epsfig,amsfonts,amssymb}
\usepackage{bm}% bold math
\usepackage{times}
\usepackage{lipsum}

\begin{document}

\title{Nonadiabatic transitions in exactly solvable quantum mechanical multichannel model: role of level curvature and counterintuitive behavior}

\author {N.~A. {Sinitsyn}$^{a}$ }
\address{$^a$ Theoretical Division, Los Alamos National Laboratory, Los Alamos, NM 87545,  USA}

\begin{abstract}
We derive an exact solution of an explicitly time-dependent multichannel model of quantum mechanical nonadiabatic transitions. In  the limit $N \gg 1$, where $N$ is the number of states, we find that the survival  probability of the initially populated state remains finite despite an almost arbitrary choice of a large number of parameters.
This observation proves that quantum mechanical nonadiabatic transitions among a large number of states can effectively keep memory about the initial state of the system. This property can lead to a strongly non-ergodic behavior even in the thermodynamic limit of some systems with a broad distribution of coupling constants and the lack of energy conservation.   
\end{abstract}

\date{\today}

\maketitle

%\section{Introduction}
 
%Explicitly time-dependent Schr\"odinger equation is
%of importance for numerous applications. 
Quantum mechanical nonadiabatic transitions have been studied in molecular collisions for long time \cite{rozen, nikitin, osherov}.
Relatively recently, multichannel nonadiabatic processes became in the focus of  research in condensed matter physics due to the interest in control of quantum many-body states in electronics, magnetic systems, and Bose condensates \cite{app-el,app-bose,app-spin,app-exp}.  
 Condensed matter physicists often achieve an insight about their systems by finding a proper model from a large variety of exactly solvable systems with well defined thermodynamic limit. Unfortunately, most of such known models correspond to the  stationary  Schr\"odinger equation. Exact solutions of explicitly time-dependent systems with a large number of states are  rare. 
 %They are also often trivially reducible to independent oscillators or non-interacting spins. 
 %It becomes important for the future progress of this field to explore new integrable models of quantum mechanical evolution with time-dependent parameters and with a large, possibly macroscopic, number of interacting states. 

The multistate version of the Stueckelberg-Majorana-Landau-Zener (LZ) model \cite{maj,LZ} is one of the most investigated explicitly time-dependent problems \cite{mlz-1,mlz-2,do,reducible,bow-tie}. Its goal is to find transition probabilities among $N$ discrete states induced during the time evolution from $\tau=-\infty$ to $\tau=+\infty$ in systems described by the Sch\"odinger equation with linearly time-dependent coefficients:
\begin{equation}
i\frac{d\psi}{d\tau} = (\hat{A} +\hat{B}\tau)\psi,
\label{mlz}
\end{equation} 
where $\hat{A}$ and $\hat{B}$ are constant $N\times N$ matrices. Matrix $\hat{B}$ is diagonal.  
Eigenstates of  $\hat{B}$ are called the {\it diabatic} states and off-diagonal elements of the matrix $\hat{A}$ are called the coupling constants.  One has to find the scattering $N\times N$ matrix $\hat{S}$, which element $S_{nn'}$ is the amplitude of the diabatic state $n'$ at $\tau \rightarrow +\infty$, given that at $\tau \rightarrow -\infty$ the system was at the state $n$.  The related matrix $\hat{P}$, $P_{nn'}=|S_{nn'}|^2$, is called the matrix of transition probabilities.
A number of exact results in which the regime $N\gg 1$ can be effectively considered, such as the Demkov-Osherov \cite{do} model, are currently known. 
%These solutions have already been applied in condensed matter physics. 

Available exact results seem to indicate that transition probabilities in the model (\ref{mlz}) behave somewhat ``classically": All known exact solutions with a finite number of states can be either reduced to independent 2-state systems \cite{reducible}, or can be completely understood in terms of individual pairwise LZ transitions that happen in a chronological order between each pair of diabatic states \cite{mlz-1,do,bow-tie}. Even beyond Eq. (\ref{mlz}), Ostrovsky studied a similar model to (\ref{mlz})  with nonlinear level crossings and came to similar conclusions \cite{coulomb}.

Numerical simulations show deviations from the semiclassical behavior in non-integrable cases  but  the available exact results and numerical simulations of the model (\ref{mlz}) with large sizes of the phase space \cite{dobrovitski}  suggest that  dynamics of  such driven multistate systems can be at least qualitatively and approximately understood in 
terms of the  successive two-state interactions that happen at moments when pairs of diabatic energies  cross or pass close to  each other. Pairwise transition amplitudes then can be used to estimate amplitudes of  the full trajectories in the phase space.
 
%%%%%%%%%%%%%%%%%%%%%%%%%%%%%%%%%%%%%%%%%%%%%%%%%%%%%%%%%%%%%%%%%%%%%%%%%%%%%%%%%%%%%%%%%%%
\begin{figure}%[!htb]
\scalebox{0.35}[0.35]{\includegraphics{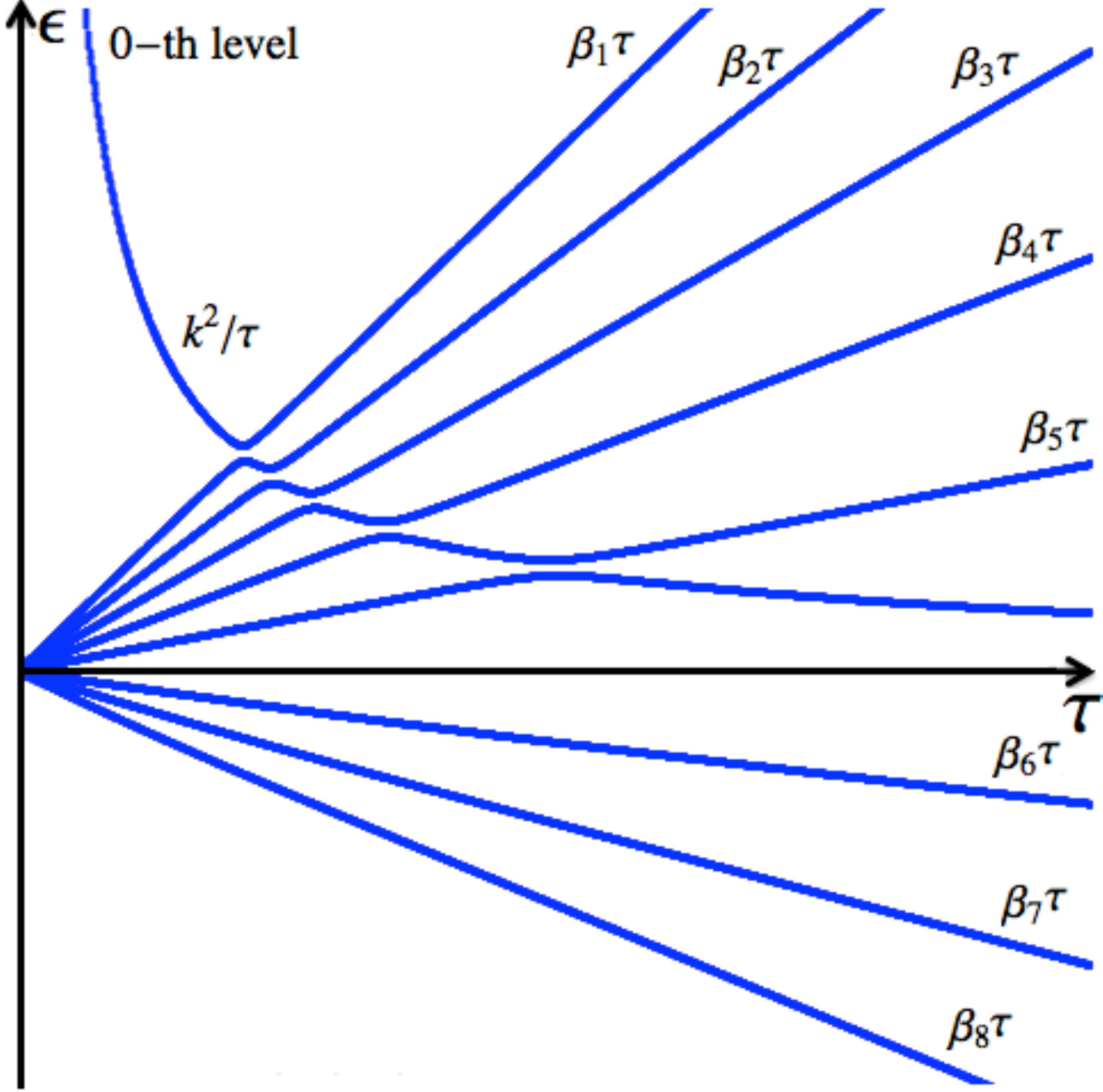}}
%\scalebox{0.17}[0.18]{\includegraphics{contour2.pdf}}
\hspace{-2mm}\vspace{-4mm}   
\caption{Time-dependence of the eigenspectrum (adiabatic energies) of the Hamiltonian with elements $H_{00}=k^2/\tau$, $H_{nn} = \beta_n \tau$, $H_{n0}=H_{0n}=g_n$, and zero otherwise ($n=1,\ldots, N$; $N=8$). %., as Hermitian Hamiltonians only have real eigenvalues.
} \label{rays}
\end{figure}
%%%%%%%%%%%%%%%%%%%%%%%%%%%%%%%%%%%%%%%%%%%%%%%%%%%%%%%%%%%%%%%%%%%%%%%%%%%%%%%%%%%%%%%%%%%
 
  Based mostly on the available exact solutions of (\ref{mlz}), it has been argued in various contexts \cite{app-exp}   that if a  level with an extremal slope crosses diabatic energies of a large number $N$ of initially unpopulated states then the probability $P_{0\rightarrow 0}$ to remain in the initial state after all intersections decreases exponentially with $N$, i.e. $P_{0\rightarrow 0} \sim {\rm exp}(-\alpha N) $, where  $\alpha$ is a coefficient, which characterizes a  transition probability at a typical encountered avoided crossing point. Such a result would be perfectly justified in classical stochastic kinetics \cite{prokofiev}.  Quantum mechanical exact results only extended the validity of such semiclassical expectations. There can certainly be exceptions that correspond to resonances that can encounter when a large number of parameters are fine-tuned but this does not change the general qualitative understanding.
 
  In this letter, we provide an explicit example, which unambiguously demonstrates
that the above intuition about the large-$N$ limit of quantum mechanical nonadiabatic transitions becomes incorrect if we consider the evolution beyond the linear time dependence of diabatic energies. We will show that even if $N$ is large and all  parameters in the Hamiltonian are chosen almost arbitrarily, e.g. from some random distribution, the probability  to remain at the initial level after all level intersections can remain $O(1)$, i.e. essentially finite.

 To achieve our goal, we will solve a model of $(N+1)$ interacting states with the evolution equation for amplitudes
 \begin{equation}
 i\frac{d}{d\tau} a=\frac{k^2}{\tau} a+\sum \limits_{n=1}^N g_n b_n, \quad
i\frac{d}{d\tau} b_n=\beta_n \tau b_n+g_n a,
\label{mmod1}
\end{equation}
where $n=1,\ldots N$.  We will call the diabatic state with the amplitude $a$ the $0$-th level.  Its diabatic energy (i.e the energy that it would have at zero coupling to other states) decays according to the Coulomb law, $\sim 1/\tau$, while diabatic energies of other 
states change linearly, as shown in Fig. \ref{rays}. For this property, we will call the model (\ref{mmod1}) the LZC-model, after Landau, Zener and Coulomb.
%This state interacts with $N$ other states having amplitudes $b_n$.
 At $\tau \rightarrow 0$, transitions among states do not happen because the $0$-th state has infinite  energy.
 % Initially, $N$ other levels  are degenerate but  their diabatic energies linearly split with time. 
 We will assume that the $0$-th level is populated at $\tau \rightarrow 0$, while other states have vanishing amplitudes.
 During the evolution, for some time, distances between diabatic energies  become comparable to coupling constants. This is the time when nonadiabatic transitions  mostly happen. At $\tau \rightarrow +\infty$ all diabatic energies become again well separated and transitions  terminate. %The scattering matrix for the evolution of $\tau$ from zero to infinity is consequently well defined.

%%%%%%%%%%%%%%%%%%%%%%%%%%%%%%%%%%%%%%%%%%%%%%%%%%%%%%%%%%%%%%%%%%%%%%%%%%%%%%%%%%%%%%%%%%%
\begin{figure}%[!htb]
\scalebox{0.17}[0.17]{\includegraphics{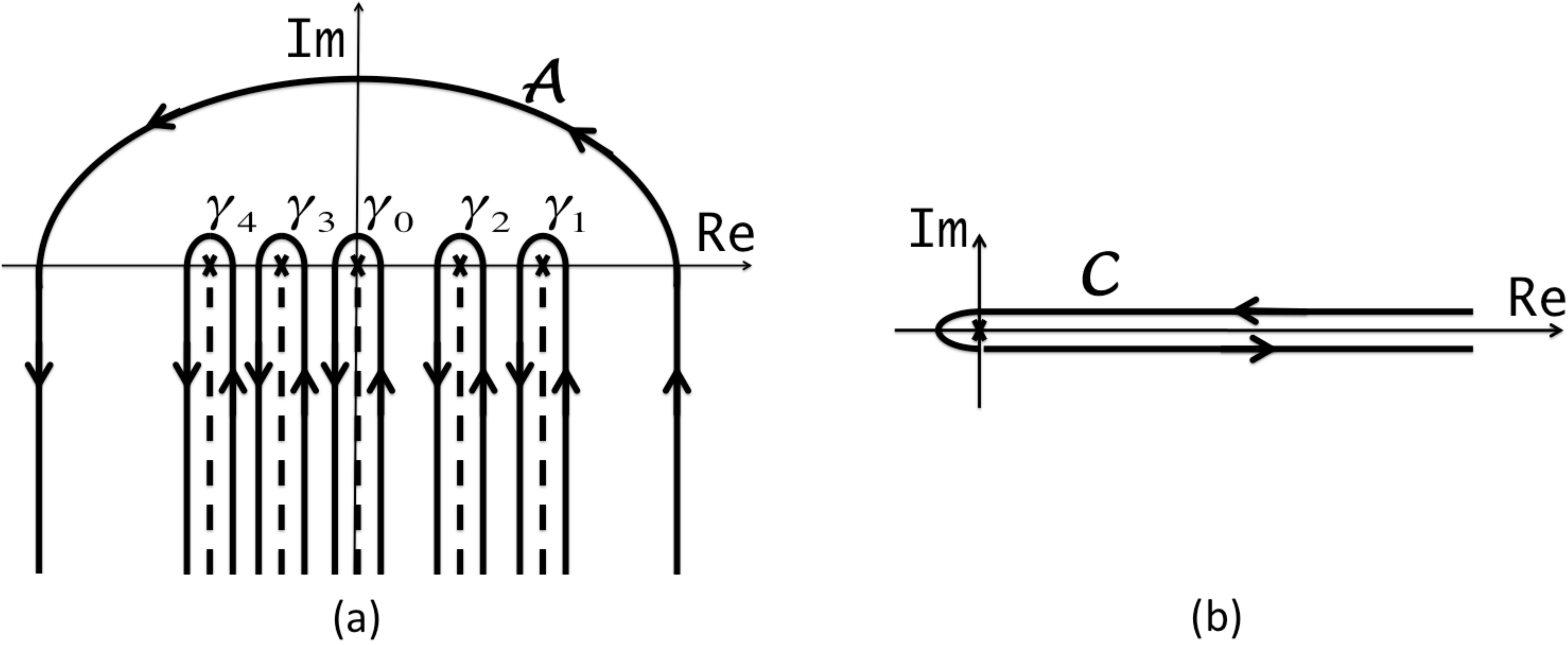}}
%\scalebox{0.17}[0.18]{\includegraphics{contour2.pdf}}
\hspace{-2mm}\vspace{-4mm}   
\caption{(a) The integration contour ${\bm A}$ incloses all branch cuts (dashed lines) from a large distance. The integral over this contour can then be expressed as the sum of the integrals over contours ${\bm \gamma_n}$, $n=0,\ldots,N$ along the branch cuts. (b) By a change of variables, each integral over ${\bm \gamma_n}$ can be transformed into the integral over the contour ${\bm C}$.
%., as Hermitian Hamiltonians only have real eigenvalues.
} \label{cont}
\end{figure}
%%%%%%%%%%%%%%%%%%%%%%%%%%%%%%%%%%%%%%%%%%%%%%%%%%%%%%%%%%%%%%%%%%%%%%%%%%%%%%%%%%%%%%%%%%%
From the semiclassical picture, we expect that the behavior of the transition probabilities will be mainly influenced by the encountered  crossing points of the $0$-th level with levels $n$ having $\beta_n>0$, as shown in Fig. \ref{rays}. Drawing the intuition from \cite{app-exp}, one may also expect that the probability to remain at the initial $0$-th level after all intersections will be exponentially suppressed at large $N$. 

We will solve the LZC model  by changing variables as $a = \tau b_0$ and $t=\tau^2/2$, leading to 
\begin{equation}
 2it \frac{db_0}{dt}  = (k^2-i) b_0+\sum_{n=1}^N g_n b_n, \quad
i\frac{d b_n}{dt} = \beta_n  b_n+g_n b_0.
\label{mmod2}
\end{equation}
We then introduce the anzatz
\begin{equation}
b_n(t)=\int_{{\bm A}} du e^{-iut} B_n(u), \quad n=0,\ldots, N,
\label{ft1}
\end{equation}
where ${\bm A}$ is a contour such that the integrand vanishes at this contour limits. Substituting (\ref{ft1}) in (\ref{mmod2}) we obtain a 1st order differential equation for $B_0(u)$, which is trivially solvable. Substituting the result back to (\ref{ft1}) we find
%\begin{widetext}
\begin{eqnarray}
\nonumber b_0(t) &=& Q\int_{{\bm A}} du \, e^{-iut} (-u)^{-\frac{1}{2} +i \frac{k^2}{2}} \prod_{n=1}^N\left( \frac{-u+\beta_n}{-u}\right)^{\frac{ig_n^2}{2\beta_n}},\\
%\nonumber
 \label{sol11} \\
\nonumber b_{j}(t) &=& -Qg_j \int_{{\bm A}} du \, \frac{ e^{-iut} (-u)^{\frac{-1+ik^2}{2} }}{-u+\beta_j} \prod_{n=1}^N \left( \frac{-u+\beta_n}{-u}\right)^{\frac{ig_n^2}{2\beta_n}}.  
\end{eqnarray} 
%\end{widetext}
Integrals in (\ref{sol11}) have almost the same structure and the same initial conditions for amplitudes as in the model of Ostrovsky \cite{coulomb}. Following his steps,
we consider a contour ${\bm A}$ that incloses branch cuts at $u=(\{ \beta_n \}, 0)$ from a large distance and goes to infinities at $u = -i\infty \pm R$, where $R$ is a large real number. In this limit, we can disregard  terms $\beta_n$ in comparison with $u$, so that integrals in (\ref{sol11}) simplify, e.g., 
\begin{equation}
b_0(t) \rightarrow Q\int_{{\bm A}} e^{-iut} (-u)^{-\frac{1}{2} +i \frac{k^2}{2} } \,d u.\\
%\nonumber \\
%b^{\bm C}_{j\ne 0}(t) &\approx & -Qg_j \int_{{\bm C}} e^{-iut} (-u)^{-\frac{3}{2} +i \frac{k^2}{2}}  
\label{sol21}
\end{equation} 
In (\ref{sol21}), the contour ${\bm A}$  can be transformed into the contour ${\bm C}$ in Fig. \ref{cont}(b) by 
switching to the variable $z=iut$ and shrinking the contour to run around the branch cut of $z$. We can then use the formula
\begin{equation}
\Gamma(z) =-\frac{1}{2i{\rm sin}(\pi z)} \int_{\bm C} (-\tau)^{z-1} e^{-\tau} d\tau
\label{GG11}
\end{equation}
to evaluate the integral.
%obtain the  in the complex $\tau$-plane starts from $+\infty$, approaches $t=0$ along the real axis, encircles this point counterclockwise and returns to $+\infty$.
Returning to original variables: $a(\tau) = \tau b_0(\tau^2/2)$ and $b_n(\tau) =b_n(\tau^2/2)$, we  find that, at $\tau=0_+$, all $b_n$ vanish, while  $a$ remains finite. If we set
\begin{equation}
Q=\frac{e^{+3i\pi/4-i{\rm arg}\Gamma(1/2+ik^2/2) -ik^2 {\rm ln}(2)/2}} {\sqrt{4\pi} \sqrt{1+e^{\pi k^2}}},
\label{qq1}
\end{equation}
then we will arrange at $\tau \rightarrow 0_+$  that 
\begin{equation}
a(\tau)_{\tau \rightarrow 0} \sim  \tau^{-ik^2}, \quad b_{j}(\tau)_{\tau \rightarrow 0}  \sim   O(\tau), \quad j \ne 0,
\label{int21}
\end{equation}
which corresponds to the initially populated $0$-th level asymptotic.
%We note here that if the population of the $0$-th level is initially equal to $1$ then at $t\rightarrow 0$ the leading asymptotic for $b^{\bm C}_n(t)$ are given by
In order to find transition probabilities at $\tau \rightarrow +\infty$ limit, we continuously deform the contour ${\bm A}$ into a combination of contours ${\bm \gamma_n}$ that inclose the branch cuts  at $u=\beta_n$ as shown in Fig. \ref{cont}(a).
In the limit $t\rightarrow +\infty$, only the vicinity of the branching points contribute essentially to each integral over ${\bm \gamma_n}$. Hence one can change variables $u \rightarrow u+\beta_n$, keeping the dependence on $u$ only for terms that are singular near the origin of the ${\bm \gamma_n}$-th cut. In all other factors, we can substitute $u$ by its value at this point. Simple dimensionality arguments show that the integration for  $b_{j}(t)_{t\rightarrow +\infty}$ in (\ref{sol11}) over ${\bm \gamma_{i}}$  remains finite only if $i=j$. Hence a $j$-th integral in (\ref{sol11}) over  ${\bm \gamma_{j}}$ provides the 
 asymptotic at $t\rightarrow +\infty$ for  $b_{j}(t)$, i.e.
 \begin{widetext}
\begin{eqnarray}
\nonumber b_0(t)_{\rightarrow+\infty} &\sim & Q \left( \prod_{j=1}^N \left(\beta_j\right)^{\frac{ig_j^2}{2\beta_j}}  \right) \int_{{\bm \gamma_0}} du \, e^{-iut} (-u)^{-\frac{1}{2} +i \frac{k^2}{2} -i\sum_{j=1}^N \frac{ig_j^2}{2\beta_j}},  \\
%\nonumber &=& 2 Q\sqrt{\pi}t^{-1/2+ik^2/2} e^{-i\pi/4+i{\rm arg}\Gamma(1/2+ik^2/2)} \sqrt{1+e^{-\pi k^2}}, \\
\label{sol44} \\
\nonumber b_{j}(t)_{\rightarrow+\infty} & \sim &  -Qg_j \left( (-\beta_j)^{-\frac{1}{2} +i\frac{k^2}{2} - i \sum_{n=1}^N \frac{g_n^2}{2\beta_n}} \prod_{n=1, \, n \ne j}^N (\beta_n - \beta_j)^{i\frac{g_n^2}{2\beta_n}} \right) \int_{\bm \gamma_0} du \, e^{-iut} (-u)^{-1+i\frac{g_j^2}{2\beta_j}}, \quad j \ne 0, 
\end{eqnarray}  
\end{widetext}
where we should assume that $(-i)=e^{-i\pi/2}$ and $-1=e^{-i\pi}$. Remaining integrals again can be evaluated with Eq. (\ref{GG11}).
% As for non-vanishing contributions, the result of integration depends on the relative values of $\beta_n$ and their sign.  Let introduce a convention that  indexes $l$ and $m$ correspond to$\beta_l>0$, and $\beta_m>0$ and $l>m$,  while indexes $s$ and $r$ correspond to levels with $\beta_r<0$, and $\beta_s<0$ and $r>s$.
In the supplementary file \cite{supplementary} we provide  explicit expressions for  asymptotic amplitudes at $\tau \rightarrow +\infty$ that we obtained from (\ref{sol44}).
Transition probabilities 
can be obtained by taking squares of the absolute values of transition amplitudes. Surprisingly, the final result appears very simple looking. 
In order to write it, it is convenient to introduce LZ-like  probabilities:
\begin{equation}
p_j = e^{-\pi g_j^2/|\beta_j|}, \quad j=1,\ldots, N.
\label{plz}
\end{equation} 
%%%%%%%%%%%%%%%%%%%%%%%%%%%%%%%%%%%%%%%%%%%%%%%%%%%%%%%%%%%%%%%%%%%%%%%%%%%%%%%%%%%%%%%%%%%
\begin{figure}%[!htb]
\scalebox{0.28}[0.28]{\includegraphics{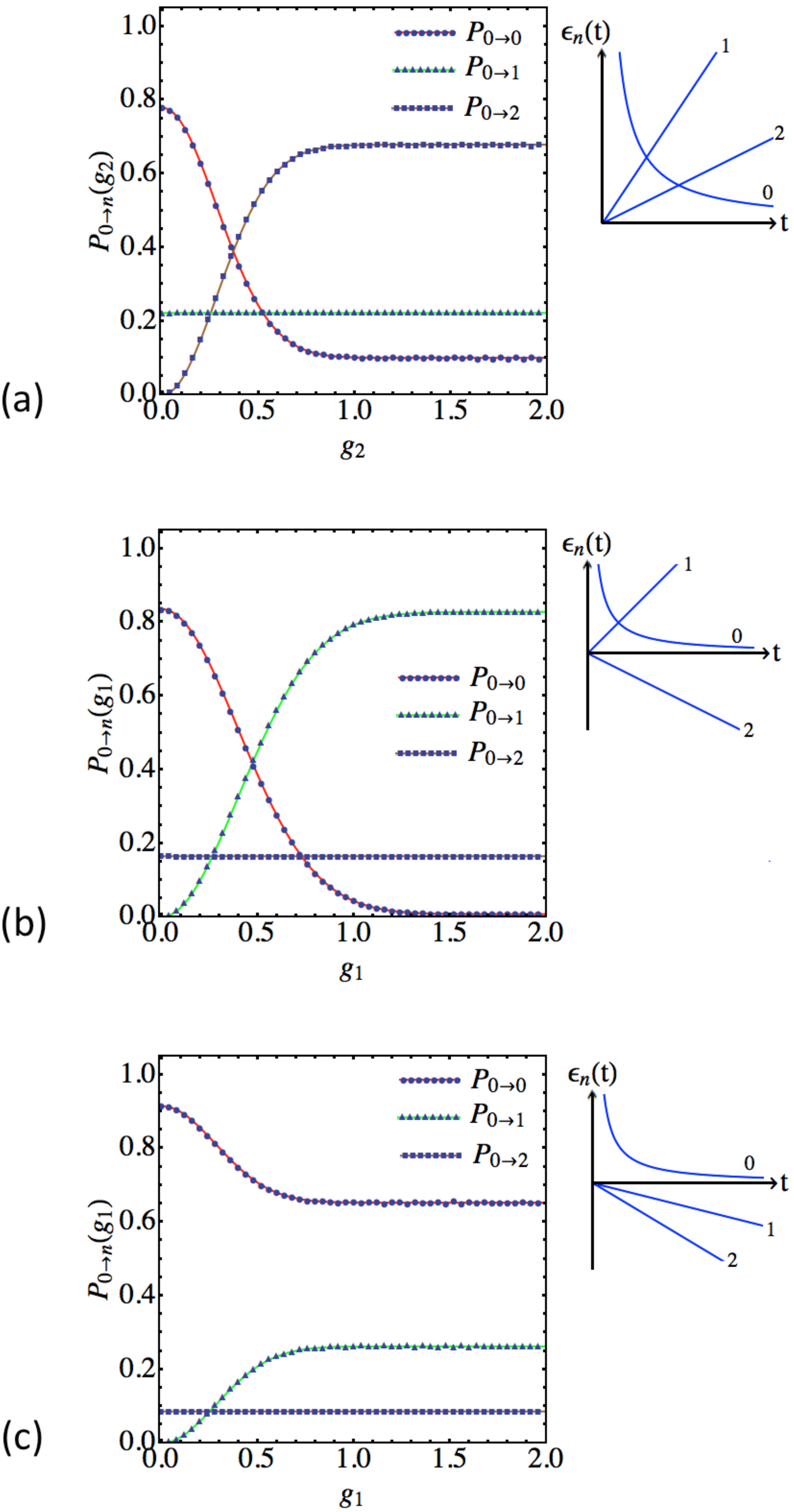}}
\hspace{-2mm}\vspace{-4mm}   
\caption{Probabilities of diabatic states at time $t\rightarrow +\infty$ in (2+1)-LZC model  for (a)  $\beta_1>\beta_2 >0$; $k^2=0.7$, $g_1=0.3$, $\beta_1=1$, $\beta_2=0.5$; (b) 
$\beta_1>0$, $\beta_2 <0$; $k^2=0.5$, $g_2=1.0$, $\beta_1=1$, $\beta_2=-1$; (c) $\beta_1<0$, $\beta_2 <0$; $k^2=0.2$, $g_2=0.3$, $\beta_1=-0.5$, $\beta_2=-1$.  
Solid curves are theoretical predictions by Eqs. (\ref{p00})-(\ref{p0j2}). Discrete points represent the results of the numerical calculations.  Insets on the right illustrate  time-dependence of diabatic energies. In all numerical tests, the time evolves from $\tau=0.0001$ to $\tau=800$. 
%., as Hermitian Hamiltonians only have real eigenvalues.
} 
\label{check}
\end{figure}
%%%%%%%%%%%%%%%%%%%%%%%%%%%%%%%%%%%%%%%%%%%%%%%%%%%%%%%%%%%%%%%%%%%%%%%%%%%%%%%%%%%%%%%%%%%
The transition probabilities from the initially populated $0$-th level to all possible states are then given by 
\begin{equation}
\noindent P_{0\rightarrow 0}= \frac{\left( \prod \limits_{n}^{\beta_n>0} p_n \right) +e^{-\pi k^2} \left( \prod \limits_{m}^{\beta_m<0} p_m \right)}{1+e^{-\pi k^2}},
\label{p00}
\end{equation}

\begin{equation}
\noindent  \underline{ \beta_j>0:} \quad P_{0\rightarrow j}= \frac{ \left( \prod \limits_{n}^{\beta_n>\beta_j} p_n \right) (1-p_j ) }{1+e^{-\pi k^2}}, 
\label{p0j1}
\end{equation}

\begin{equation}
\noindent  \underline{ \beta_j<0:} \quad P_{0\rightarrow j}= \frac{ \left( \prod \limits_{n}^{\beta_n<\beta_j} p_n \right) (1-p_j ) e^{-\pi k^2} }{1+e^{-\pi k^2}}.
\label{p0j2}
\end{equation}
%At first view, Eqs. (\ref{p00})-(\ref{p0j2}) are quite simple. 
Here, for example,  $\prod \limits_{n}^{\beta_n>0}p_n $ means the product over all $p_n$ with indexes $n$ of states having slopes $\beta_n$ larger than zero. If there are no such states, this product is set to unity.
Several qualitative observations  follow straightforwardly:

\underline{Corollary  1:}  The transition probability from the $0$-th level to the level $j$   with $\beta_j>0$  depends only on coupling constants and slopes of the levels $n$ with $\beta_n \ge \beta_j$.  

\underline{Corollary  2:}  The transition probability from the $0$-th level to the level $j$   with $\beta_j<0$  depends only on coupling constants and slopes of the levels $n$ with $\beta_n \le \beta_j$.

%%%%%%%%%%%%%%%%%%%%%%%%%%%%%%%%%%%%%%%%%%%%%%%%%%%%%%%%%%%%%%%%%%%%%%%%%%%%%%%%%%%%%%%%%%%
%\begin{figure}%[!htb]
%\scalebox{0.275}[0.275]{\includegraphics{check5.pdf}}
%\hspace{-2mm}\vspace{-4mm}   
%\caption{Probabilities of diabatic states at time $t\rightarrow +\infty$  and  at $\beta_1>0$, $\beta_2 <0$ as functions of the coupling $g_1$ to level-1.  Parameters: $k^2=0.5$, $g_2=1.0$, $\beta_1=1$, $\beta_2=-1$.
%}
 %\label{cross}
%\end{figure}
%%%%%%%%%%%%%%%%%%%%%%%%%%%%%%%%%%%%%%%%%%%%%%%%%%%%%%%%%%%%%%%%%%%%%%%%%%%%%%%%%%%%%%%%%%%

Corollary 1 would be  expected from the semiclassical point of view. 
In the independent crossing approximation, only crossing points that precede the crossing of the level $j$ would influence the population of the $j$-th level. 
This situation has previously encountered in the solution of the Demkov-Osherov model \cite{do}. Parameter $k$ characterizes the curvature of the $0$-th level near the avoided crossings \cite{supplementary}. 
The limit of large values of $k$ corresponds to locally linear dependence of $0$-th diabatic energy on time, so that the formulas (\ref{p00})-(\ref{p0j1}) naturally reproduce  the results of the Demkov-Osherov formula \cite{do} if we set $k\rightarrow +\infty$.    

In contrast,  Corollary 2 is very counterintuitive. Surprisingly, none of  the encountered avoided crossing points influences transition probabilities to the levels $j$ with $\beta_j <0$, no matter how strong can couplings at crossing poins be. Moreover, only couplings to the levels $n$
with $\beta_n<\beta_j$, i.e. the levels that diverge faster than the level $j$ from the $0$-th level, contribute to the transition  probability to the $j$-th diabatic state. 

In Fig. \ref{check}, we compare our theoretical predictions (\ref{p00})-(\ref{p0j2}) with transition probabilities obtained in our numerical studies of the evolution (\ref{mmod2}) in a three-state LZC model. We find an excellent agreement 
between numerical and theoretical results. In particular, Fig. \ref{check}(a) shows that if $\beta_1 > \beta_2>0$ then the transition probability to the level with the slope $\beta_1$ does not depend on the coupling $g_2$ to the level with a lower slope,  in agreement with Corollary 1. In case when  $\beta_1 > \beta_2$ and simultaneously $\beta_2<0$, we find that already the transition probability to the level with the slope $\beta_2$ does not depend on the coupling $g_1$ between the $0$-th level and the level with the slope $\beta_1$. 

\underline{Corollary  3:}  The probability to stay at the  $0$-th level at $\tau \rightarrow +\infty$ remains finite if either $\beta_j>0$ or $\beta_j<0$ for all $j=1,\ldots N$.  Moreover, in the limit $N \rightarrow \infty$ and at finite couplings, the probability to remain 
in the $0$-th level depends only on the parameter $k$.

The last observation is the main result of this letter. It is normally expected that if a $0$-th level interacts with a large number of states with arbitrary coupling strengths, such an interaction would lead to a sort of equilibration. It has been often argued that the survival probability should be, actually, vanishing exponentially with the number of levels $N$ that interact with the highest slope level \cite{app-exp}. The $0$-th level satisfies this condition if conditions of Corollary 3 are satisfied.   Nevertheless,  Eq. (\ref{p00}) predicts that if all $\beta_j >0$ then $P_{0\rightarrow 0} > e^{-\pi k^2}/(1+e^{-\pi k^2})$, and if all $\beta_j <0$ then $P_{0\rightarrow 0} > 1/(1+e^{-\pi k^2})$.

 %Interestingly, also, the sum of these minimal values of $P_{0\rightarrow 0}$ add to unity. 

%%%%%%%%%%%%%%%%%%%%%%%%%%%%%%%%%%%%%%%%%%%%%%%%%%%%%%%%%%%%%%%%%%%%%%%%%%%%%%%%%%%%%%%%%%%
%\begin{figure}%[!htb]
%\scalebox{0.275}[0.275]{\includegraphics{check6.pdf}}
%\hspace{-2mm}\vspace{-4mm}   
%\caption{Probabilities of diabatic states at time $t\rightarrow +\infty$ , $\beta_1<0$ and $\beta_2 <0$, as functions of the coupling $g_1$ to level-1.  Parameters: $k^2=0.2$, $g_2=0.3$, $\beta_1=-0.5$, $\beta_2=-1$.
%}
 %\label{cross}
%\end{figure}
%%%%%%%%%%%%%%%%%%%%%%%%%%%%%%%%%%%%%%%%%%%%%%%%%%%%%%%%%%%%%%%%%%%%%%%%%%%%%%%%%%%%%%%%%%%

{\it In conclusion,} we determined transition probabilities in a multichannel model of nonadiabatic transitions with a  well defined transition probability matrix among all diabatic states.
In our model, one of the diabatic energy levels changed nonlinearly, which allowed us to investigate the effect of the curvature of this level on the transition probability matrix.
We found that the behavior of the probability to remain at the initially populated level, as well as the levels having a negative slope of diabatic energy, can behave very counterintuitively. For example,
some of the transition probabilities can be finite but independent of the couplings at {\it all} avoided crossing points that encounter during the system's evolution.   
%The prediction of an exponential decay of the initially populated level in the limit of a large number $N$ of states coupled to it  has been made for many driven mesoscopic systems \cite{app-exp}. While those predictions remain valid for linear level crossings, our results show that generally the probability to stay at the initial diabatic state can remain finite no matter how large are the couplings and what is their distribution. In contrast to previous expectations,

We predict that systems with nonadiabatic transitions in the large $N$ limit can show strongly non-classical behavior, e.g. with some states keeping a finite amplitude  despite a large distribution of coupling constants.
%Our example shows that the standard approach to introduce environment by calculating the 2nd order response of each of its degrees of freedom and then finding the effective evolution equation for the density matrix of the central system is generally insufficient even at  the large $N$ limit with a broad randomness of coupling constants and other parameters.
In supplementary file \cite{supplementary}, we demonstrate numerically that this property appears also in a wide class of  diabatic potentials, e.g. of the form $ \sim 1/t^r$ with some constant parameter $r$. The independent crossing approximation works for them only when $P_{0 \rightarrow 0}$ remains close to unity, and the exponentially strong enhancement of the probability to stay on the initial level is observed otherwise. 

The Coulomb potential frequently appears in studies of nonadiabatic transitions in molecular physics \cite{rozen,coulomb,singular}. Moreover, if one of the levels is degenerate before a molecular collision or an application of a time-dependent field,  this degeneracy  experiences fan-like spitting, as in Rydberg's atoms due to the Stark effect \cite{stark}. Hence, the LZC-model represents an important class of processes in atomic and molecular physics.

%Our exact result provides only a proof that such a behavior is, in principle, possible. Future numerical and theoretical studies should determine conditions for observation of this effect, e.g. in Rydberg atoms. Numerical simulations of nonadiabatic transitions at large $N$ are challenging but possible to produce in order to study this effect in non-integrable systems. 
 
Among other applications, we mention that, in the limit $k \rightarrow 0$, the LZC-model corresponds to the previously unstudied case of an N-state bow-tie model \cite{bow-tie} with the time evolution from $\tau=0$ to $\tau=+\infty$. In the supplementary file \cite{supplementary}, we  show that the two-state version of the LZC model provides, at a little cost for complexity, a better approximation of the transition probability at an avoided crossing than the Stueckelberg-Majorana-Landau-Zener formula. It also provides a sample of a solvable model of an avoided diabatic crossing with a constant off-diagonal coupling. Finally, we provide an example in \cite{supplementary} of a possible realization of a 3-state LZC-model in an interacting qubit system.

%\newpage
%\appendix

\section*{Supplementary material}
In support of the main text, section 1 of this supplementary material
provides the explicit form of the asymptotic values of
amplitudes. Section 2 explains details of the numerical algorithm and provides additional numerical proof of the theoretical formulas in five-state LZC models. 
Section 3 discusses similar counterintuitive behavior in nonintegrable models with a singular potential of 0-th level.
Section 4 shows an example of the two-qubit Hamiltonian that realizes the three-state version of the LZC-model. 
Section 5 explores the 2-state
version of the LZC model and its applications as an alternative to the LZ-formula and as a
model for a diabatic avoided level crossing.

\date{\today}

\maketitle

\begin{widetext}
\section{Transition Amplitudes in LZC-model}

Asymptotically at  $\tau \rightarrow 0$, we choose amplitudes so that the $0$-th level is initially populated. We fix constant phases so that

\begin{eqnarray}
a(\tau)_{\tau \rightarrow 0} \sim  \tau^{-ik^2}, \quad b_{j}(\tau)_{\tau \rightarrow 0}  \sim   O(\tau), \quad j =1,\ldots N.
%\nonumber \\
%
\label{sol6}
\end{eqnarray} 

The amplitudes at $\tau \rightarrow +\infty$ are then given by

 \begin{eqnarray}
a(\tau)_{\tau \rightarrow+\infty}  & \sim &   \tau^{i \left(-k^2 +\sum_{n=1}^N \frac{\pi g_n^2}{\beta_n} \right)}  e^{i\Phi_0}  \left (\frac{e^{-\sum \limits_j^{\beta_j>0} \frac{\pi g_j^2}{\beta_j}}+e^{-\pi k^2 -\sum \limits_{j}^{\beta_j<0}  \frac{\pi g_j^2}{|\beta_j|} }}{1+e^{-\pi k^2}} \right)^{1/2}, \\
 \nonumber \\
 \underline{ \beta_j>0:} && \quad b_j(\tau)_{\tau \rightarrow+\infty}   \sim e^{-\frac{i\beta_j \tau^2}{2}} \tau^{-i\frac{\pi g_j^2}{\beta_j}} e^{i\Phi_{j}}  \left (\frac{\left(1-e^{-\frac{\pi g_j^2}{\beta_j}} \right) e^{-\sum \limits_{n}^{\beta_n>\beta_j} \frac{\pi g_n^2}{\beta_n} }}{1+e^{-\pi k^2}} \right)^{1/2},\\
 \nonumber \\
 \underline{ \beta_j<0:} && \quad b_j(\tau)_{\tau \rightarrow+\infty}   \sim e^{-\frac{i\beta_j \tau^2}{2}} \tau^{-i\frac{\pi g_j^2}{\beta_j}} e^{i\Phi_{j}}  \left (\frac{\left(1-e^{-\frac{\pi g_j^2}{|\beta_j|}} \right) e^{-\pi k^2-\sum \limits_{n}^{\beta_n<\beta_j} \frac{\pi g_n^2}{|\beta_n|} }}{1+e^{-\pi k^2}} \right)^{1/2},
\label{sol5}
\end{eqnarray}  
where
\begin{equation}
\Phi_0 = {\rm arg}\Gamma \left(\frac{1}{2} +i \frac{k^2}{2} -  i \sum_{n=1}^N \frac{g_n^2}{2\beta_n} \right) - {\rm arg} \Gamma \left( \frac{1}{2} + i\frac{k^2}{2}  \right) +  \sum_{n=1}^N \frac{g_n^2 {\rm ln}(|\beta_n|/2)}{ 2\beta_{n}},
\label{phi0}
\end{equation}
\begin{equation}
\Phi_j =\frac{\pi}{4}+ {\rm arg}\Gamma \left(\frac{ig_j^2}{2\beta_j} \right) - {\rm arg} \Gamma \left( \frac{1}{2} +i \frac{k^2}{2}  \right) + \sum_{n=1}^{N,\, n \ne j} \frac{g_n^2 {\rm ln}(|\beta_n-\beta_j|)}{ 2\beta_{n}}
 +\left(\frac{k^2}{2} - \frac{ig_j^2}{2\beta_j} \right) {\rm ln}(|\beta_j|/2).
\label{phi01}
\end{equation}

%\begin{equation}
%\Phi_j^{<} =\frac{\pi}{4}- {\rm arg}\Gamma \left(\frac{ig_j^2}{2|\beta_j|} \right) - {\rm arg} \Gamma \left( \frac{1}{2} +i \frac{k^2}{2}  \right) + \sum_{n=1}^{N,\, n \ne j} \frac{g_n^2 {\rm ln}(|\beta_n-\beta_j|)}{ 2\beta_{n}}
 %+\left(\frac{k^2}{2} - \frac{ig_j^2}{2\beta_j} \right) {\rm ln}(|\beta_j|/2)
%\label{phi02}
%\end{equation}
\end{widetext}

%%%%%%%%%%%%%%%%%%%%%%%%%%%%%%%%%%%%%%%%%%%%%%%%%%%%%%%%%%%%%%%%%%%%%%%%%%%%%%%%%%%%%%%%%%%
\begin{figure}[!htb]
\scalebox{0.29}[0.29]{\includegraphics{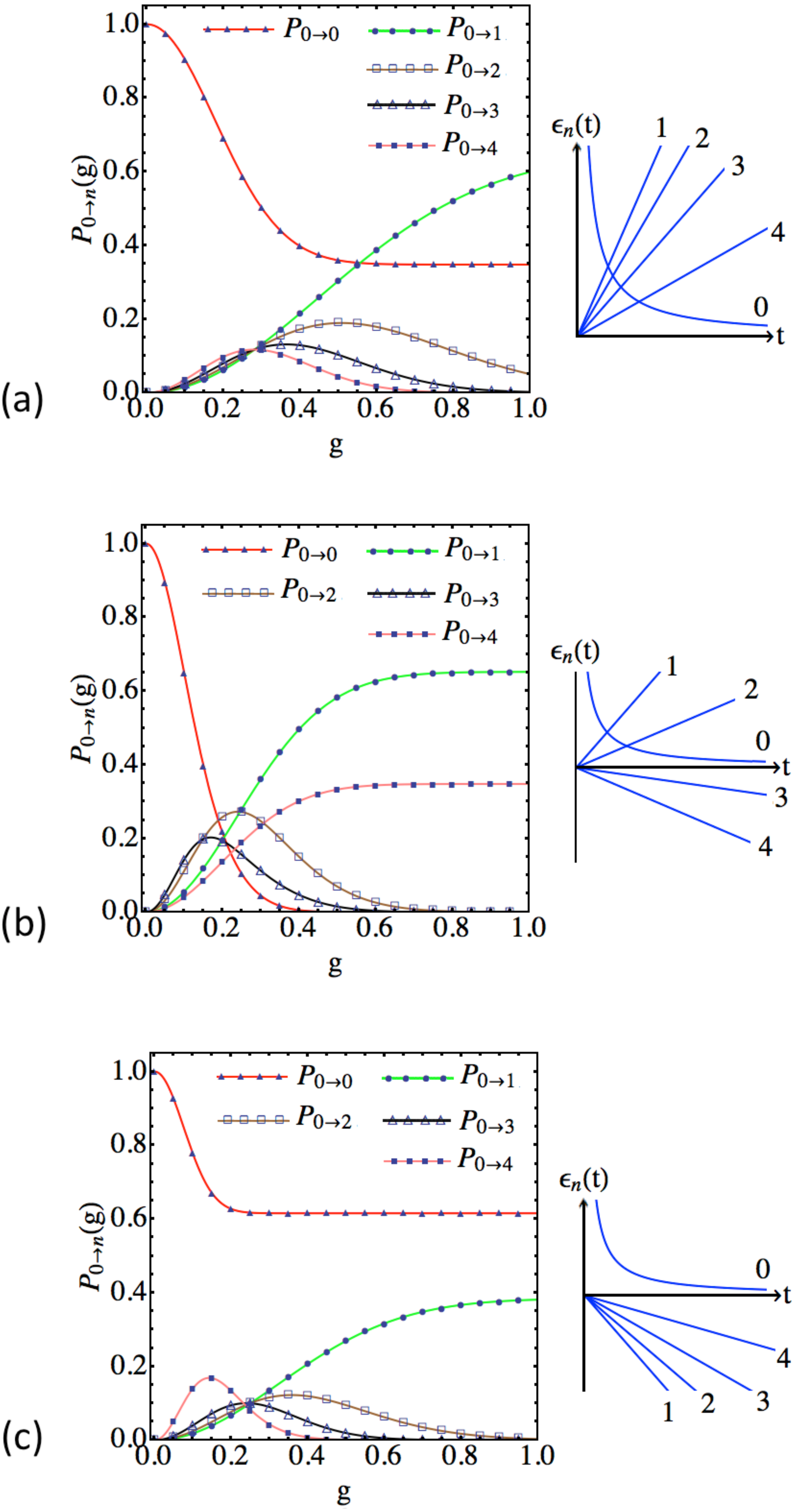}}
\hspace{-2mm}\vspace{-4mm}   
\caption{Numerical check of the transition probability formulas  in (4+1)-LZC model  for the case of equal coupling constants: $g_{j}=g$, and slopes given by  $\beta_j = b_2 - b_1*(j - 2)$ for $j \in \{ 1,2,3,4 \}$, and  
(a)  $k^2=0.2$, $b_1=0.25$, $b_2=1.25$;
(b) $k^2=0.2$, $b_1=-0.2$, $b_2=-0.65$;
(c)  $k^2=0.15$,  $\beta_1=-0.5$, $\beta_2=-1$.  
Solid curves are theoretical predictions. Discrete points represent the results of the numerical calculations.  Insets on the right illustrate  time-dependence of diabatic energies. In all numerical tests, the time evolves from $t=0.00001$ to $t=800$ with time step $dt=0.00001$. 
%., as Hermitian Hamiltonians only have real eigenvalues.
} 
\label{Fig5}
\end{figure}
%%%%%%%%%%%%%%%%%%%%%%%%%%%%%%%%%%%%%%%%%%%%%%%%

\section{Numerical tests}

In order to check our theoretical predictions, we wrote a numerical program that simulated the quantum mechanical evolution that starts at a small initial time moment $t=dt$ and proceeds in discrete time steps of sizes equal to $dt$.
The evolution at each time step is described by the evolution operator 
\begin{equation}
\hat{U}(t) = \left(\hat{1} + i\hat{H}(t) dt/2\right) \left(\hat{1} - i\hat{H}(t) dt/2\right)^{-1},
\label{ev1}
\end{equation}
where $\hat{H}(t)$ is the time-dependent matrix Hamiltonian, and $\hat{1}$ is the unit matrix. 
Operator $\hat{U}(t)$ is unitary. Up to terms $o(dt^2)$, it coincides with the true evolution operator, $\hat{\rm T} \exp \left(i\int_{t}^{t+dt} \hat{H}(t) \, dt \right)$, where $\hat{\rm T}$ is the time-ordering operator.
Being equivalent to the true evolution operator in the $dt \rightarrow 0$ limit, operator $\hat{U}(t)$ is relatively easy to calculate numerically, because the hardest step is the calculation  of the inverse of the matrix  $\left(\hat{1} - i\hat{H}(t) dt/2\right)$. By acting with $\hat{U}(t)$ on the state vector, the latter was updated up to large times at which no  changes of the state probabilities was observed up to two first significant digits.

Figure~\ref{Fig5} shows results of numerical simulations (discrete points)  versus theoretical predictions (solid curves) for a 5-state LZC-model. In this case, all coupling constants are assumed to have the same value, and probabilities were plotted as functions of this size of all couplings. Results appear to be in excellent agreement with the theoretical prediction. Figures~\ref{Fig5}(a,c) clearly show that the probability to remain in the initial level saturates at a nonzero value, which is independent of the coupling constants, as far as couplings are sufficiently large and all but 0-th level slopes have the same sign.

\section{Counterintuitive behavior in non-integrable models} 

\begin{figure}[!htb]
\scalebox{0.29}[0.29]{\includegraphics{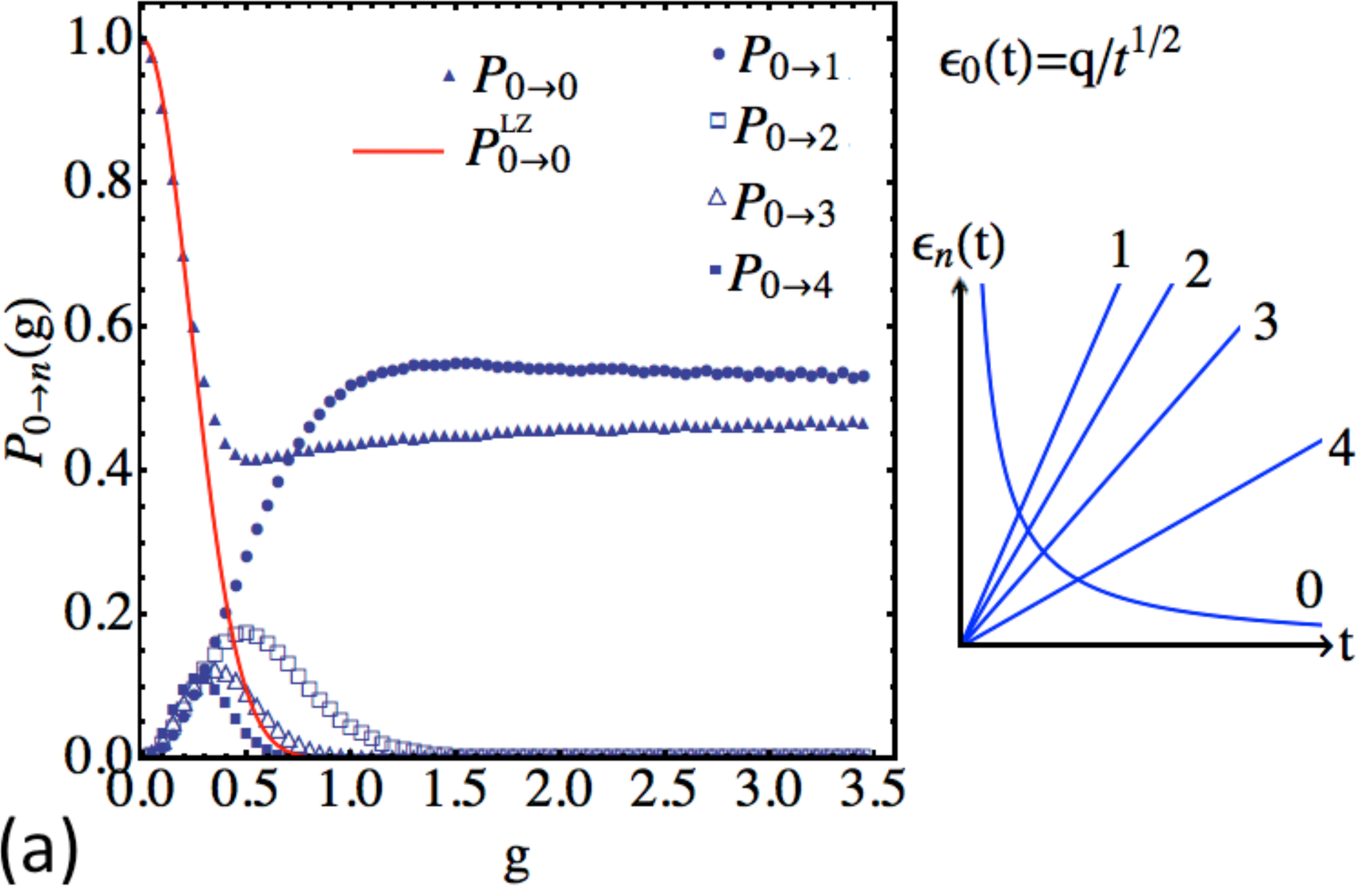}}
\scalebox{0.29}[0.29]{\includegraphics{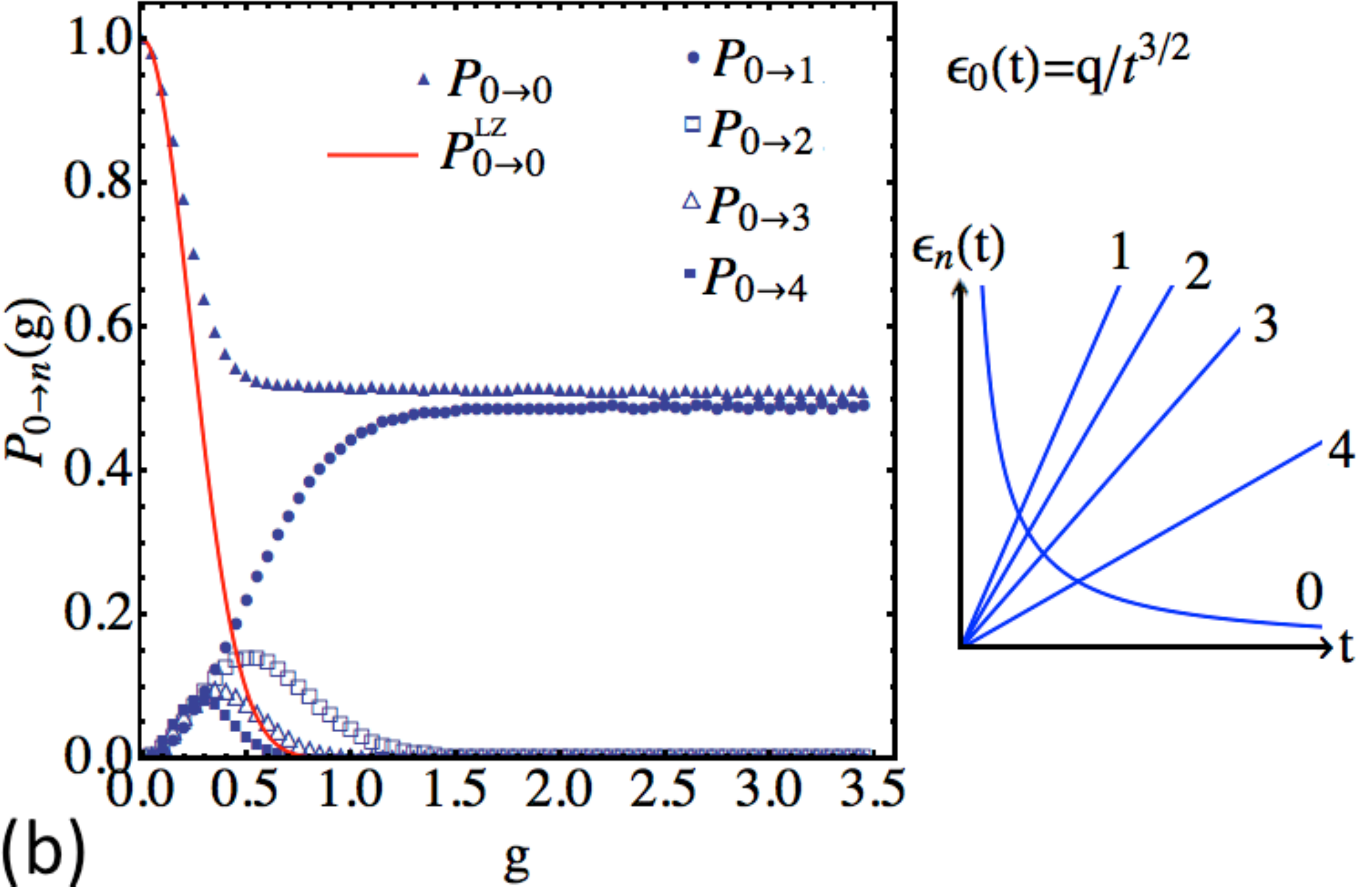}}
\scalebox{0.29}[0.29]{\includegraphics{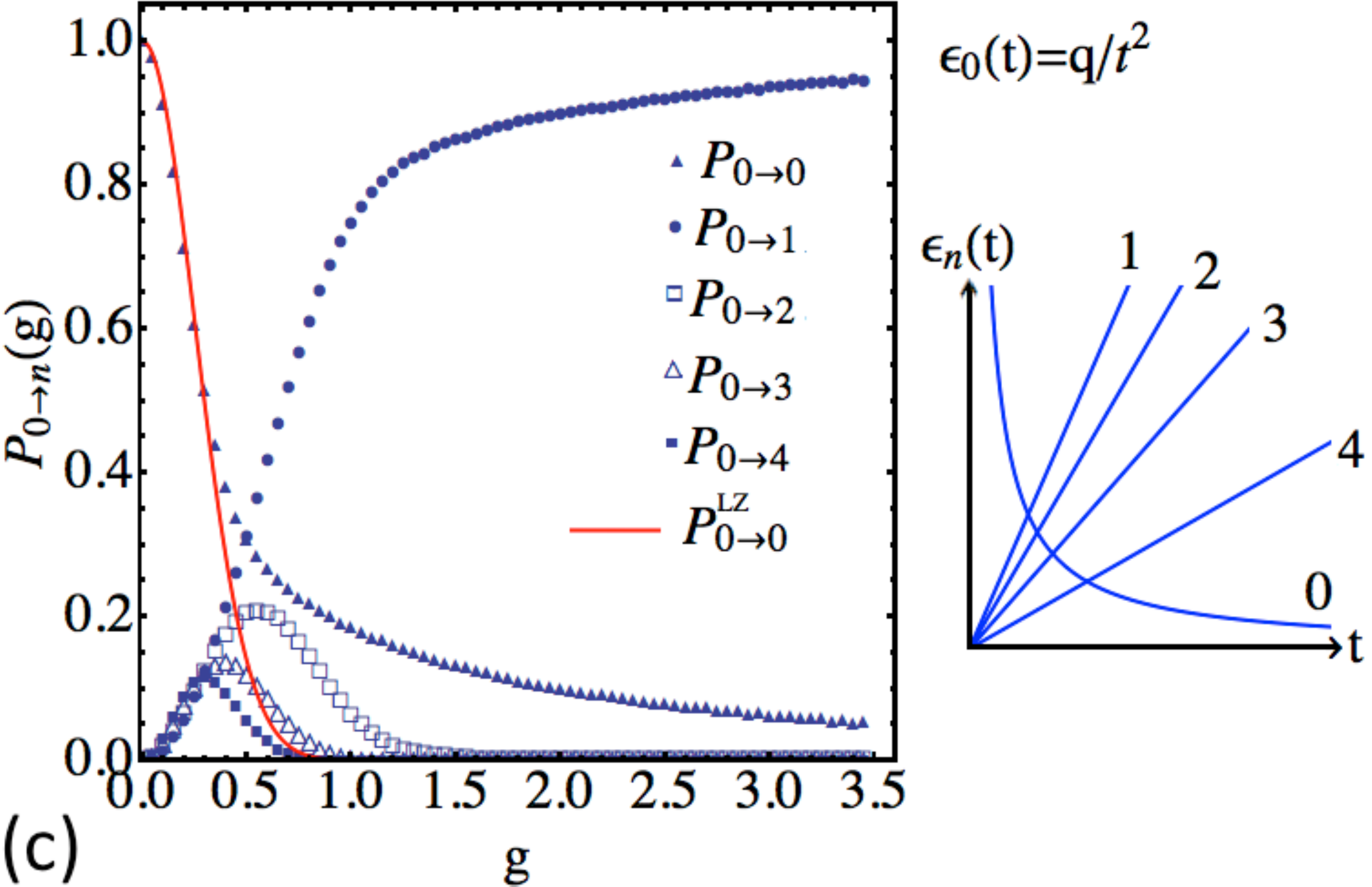}}
\hspace{-2mm}\vspace{-4mm}   
\caption{Numerically obtained transition probabilities  in a (4+1)-state model with power-law decay of the diabatic energy of the $0$-th level:  $\epsilon_0(t) = q/t^{r}$,  for $q=0.2$ (a) r=1/2, (b) r=3/2, (c) r=2. 
All coupling constants are equal: $g_{j}=g$, and slopes are given by $\beta_j = b_2 - b_1*(j - 2)$ for $j \in \{ 1,2,3,4 \}$, and   $b_1=0.25$, $b_2=1.25$. Red curve is the semiclassical prediction  of Eq.~(\ref{sem}) based on the independent crossing approximation.
%., as Hermitian Hamiltonians only have real eigenvalues.
} 
\label{quad}
\end{figure}

In this section, we explore the probability to stay at the initially occupied level with non-Coulomb potentials. 
We considered a set of diabatic potentials of the $0$-th level of the form
\begin{equation}
\epsilon_0(t) = q/t^r,
\label{diab10}
\end{equation}
with some constant parameters $q$ and $r$.  In particular, for our tests, we chose  5-state models with $q=0.2$ and $r \in \{ 2,3/2,1/2 \}$.
All other parameters were chosen as in Fig.~\ref{Fig5}(a), i.e. for the case when all other states have positive slopes of the diabatic energies. 
 Our results for transition probabilities are summarized in Fig.~\ref{quad}.

 %%%%%%%%%%%%%%%%%%%%%%%%%%%%%%%%%%%%%%%%%%%%%%%%%%%%%%%%%%%%%%%%%%%%%%%%%%%%%%%%%%%%%%%%%%%
%%%%%%%%%%%%%%%%%%%%%%%%%%%%%%%%%%%%%%%%%%%%%%%%
 %%%%%%%%%%%%%%%%%%%%%%%%%%%%%%%%%%%%%%%%%%%%%%%%%%%%%%%%%%%%%%%%%%%%%%%%%%%%%%%%%%%%%%%%%%%
%\begin{figure}[!htb]
%\scalebox{0.29}[0.29]{\includegraphics{Non2.pdf}}
%\hspace{-2mm}\vspace{-4mm}   
%\caption{Numerically obtained transition probabilities  in a (4+1)-state model with quadratic decay of the diabatic energy of the $0$-th level:  $\epsilon_0(t) = q/t^2$, $q=0.2$. All coupling constants are equal: $g_{j}=g$, and slopes are given by a formula $\beta_j = b_2 - b_1*(j - 2)$ for $j \in \{ 1,2,3,4 \}$, and   $b_1=0.25$, $b_2=1.25$.
%., as Hermitian Hamiltonians only have real eigenvalues.
%} 
%\label{quad}
%\end{figure}
%%%%%%%%%%%%%%%%%%%%%%%%%%%%%%%%%%%%%%%%%%%%%%%%

Figure~\ref{quad} shows that the phenomenon of saturation of the probability $P_{0\rightarrow 0}$ at large couplings is not an artifact of the integrability of the model. For example, Fig~\ref{quad}(a) shows that for the decay of the 0-th diabatic energy as $1/\sqrt{t}$ in a five-state model, the survival probability reaches a minimum at some coupling $g$ between states, and then it is actually growing. Figure~\ref{quad}(b) shows that the decay of $0$-th level as
$1/t^{3/2}$ produces qualitatively similar behavior to the integrable case at $r=1$. 
 For $r=2$, the probability to stay at the $0$-th level, in principle, slowly decays with increasing $g$ but this decay is not comparable to what one would expect from the independent crossing approximation. Semiclassically, one would expect that the transition probability should be close to the one predicted by
the independent crossing approximation: 
\begin{equation}
P_{0\rightarrow 0}^{LZ} \sim e^{-2\pi \sum_{i=1}^N \frac{g^2}{| b_0^i - \beta_i |}},
\label{sem}
\end{equation}
where $b_0^i=-r\beta_i$ is the slope of the $0$-th level near the intersection of its diabatic energy with the diabatic energy of the $i$-th level. The red curve in Fig.~\ref{quad} shows the prediction of the approximation (\ref{sem}). This approximation works relatively well up to $g=0.4$ but then it hopelessly fails. For example, for $g=3.5$ and $r=2$, Eq. (\ref{sem}) predicts $P_{0\rightarrow 0}^{LZ}\sim 10^{-43}$, i.e. even for such a moderate coupling strength, the difference between the prediction of Eq.~(\ref{sem}) and the numerically obtained result is over $40$ orders of magnitude.

\section{LZC Hamiltonian in the model of coupled qubits}

One may wonder if the LZC-model with more than two states can be realized experimentally. Here we show that a 3-state LZC-model is actually found in the fundamental system of two coupled qubits. The latter is the minimal hardware block needed to build a quantum computer, which explains why it was realized in a number of solid state structures and in  atomic gases. Our exactly solvable model can serve as a test for the accuracy of a control over such device structures.  

Our model is described by the Hamiltonian:
\begin{equation}
\hat{H} =J(t) \left(1 -  {\bf {\hat \sigma}^1 \cdot {\hat \sigma}^2} \right) + B_z (t) \left( {\hat \sigma}^1_z +{\hat \sigma}_z^2 \right) + B_x \left(  {\hat \sigma}^1_x -{\hat \sigma}^2_x \right),
\label{hqubit}
\end{equation}
where $\sigma^i_{\alpha}$ is the $\alpha$-component Pauli operator acting in the space of the $i$-th qubit. First term in (\ref{hqubit}) describes the Heisenberg-like coupling between qubits. Second and third terms describe  magnetic-field-like couplings.  These parameters are usually required to be controllable, i.e. time dependent on demand, for quantum information processing. 

Hamiltonian (\ref{hqubit}) would correspond to the LZC model if the following choice of parameters is assumed: 
\begin{equation}
J(t) = -\frac{k^2}{2t}, \quad B_z(t) =\frac{ \beta t}{2}, \quad B_x =-\frac{g}{ \sqrt{2} },
\label{timed}
\end{equation}  
with constants $k$ and $g$. Here we note that one does not have to reproduce singularity at $t=0$ because at sufficiently strongly coupling $J$, transitions among states will be suppressed, so it is sufficient to keep the qubit system at the ground state with large $J$ for $t<0$ and then continuously decouple them.
In this case, 
the state $\frac{1}{\sqrt{2}} \left( |\uparrow \downarrow \rangle + | \downarrow \uparrow \rangle \right) $ decouples from other three states: 
\begin{eqnarray}
\label{states-0}
|0 \rangle &=& \frac{1}{\sqrt{2}} \left( |\uparrow \downarrow \rangle - | \downarrow \uparrow \rangle \right),  \\
\label{states-1}
\nonumber \\
|1 \rangle &=&  |\uparrow \uparrow \rangle ,  \\
\nonumber \\
|2 \rangle &=&-  |\downarrow \downarrow \rangle .
\label{states}
\end{eqnarray}
In the basis (\ref{states-0})-(\ref{states}) the Hamiltonian (\ref{hqubit}) has the following matrix form:
\begin{equation}
\hat{H} = \left( 
\begin{array}{ccc}
-k^2/t & g & g \\
g & \beta t & 0 \\
g & 0 & -\beta t
\end{array}
\right),
\label{h333}
\end{equation}
which corresponds to a 3-state LZC-model.

\begin{widetext}

\section{Two-state LZC model}

\subsection{Relation to confluent hypergeometric equation}

The two-state LZC model,
\begin{equation}
i\frac{d}{d\tau}a=\frac{k^2}{\tau} a+gb,\quad
i\frac{d}{d\tau} b=\beta \tau b+ga,
\label{mod1}
\end{equation}
is trivial to solve.  Here we discuss some of the possible its applications, which to some degree, can be extended to multi-state LZC.

By changing variables $a\rightarrow \tau a'$ and $t=\tau^2 e^{-i\pi/2} \beta$, then 
%\begin{equation}
% 2it \frac{d}{dt} a=(k^2-i) a+gb, \quad
%i\frac{d}{dt} (t b)=\beta  b+ga
%\label{mod2}
%\end{equation}
differentiating 2nd equation in (\ref{mod1}) over time, we find the confluent hypergeometric equation:
\begin{equation}
t \frac{d^2}{d t^2} b+(r-t) b +s b=0, 
\label{chge}
\end{equation} 
where

\begin{equation}
r=\frac{1}{2}+i\frac{k^2}{2}, \quad s= \frac{1}{2} -i \left(\frac{g^2}{2\beta} - \frac{k^2}{2} \right)
\label{rs}
\end{equation}
are constants.
The solution of (\ref{chge}) that satisfies initial conditions $b(0)=1$, $a(0)=0$ is given by the Kramer's function $M(s,r,\tau)$. This solution has asymptotics 
\begin{equation}
M(s,r,0)=1, \quad M(s,r, \tau \rightarrow \infty) \sim \frac{\Gamma(r) e^{-\pi g^2/4\beta+if(\tau)}  }{\Gamma(s)},
\label{as}
\end{equation}
where $f(\tau)$ is a time-dependent phase factor that does not influence the transition probability. Using that $|\Gamma(\frac{1}{2}+ix)|^2=\pi/{\rm cosh}(\pi x)$, we obtain the probability to stay in the initial level:
\begin{equation}
\beta>0: \quad P_{0\rightarrow 0} =\quad P_{1\rightarrow 1}=\frac{e^{-\pi g^2/\beta} + e^{-\pi k^2}}{1+e^{-\pi k^2}},
\label{bp}
\end{equation}
\begin{equation}
\beta < 0: \quad P_{0\rightarrow 0} =\quad P_{1\rightarrow 1}= \frac{1+e^{-\pi (k^2+g^2/|\beta|) } }{1+e^{-\pi k^2}}.
\label{bm}
\end{equation}

\subsection{Alternative to LZ-formula}

Equation (\ref{bp}) takes into  account the  nonlinearity of the crossing curves.  Hence it can be used as a better approximation than the LZ-formula. Consider,  for example, an arbitrary model of a two-state avoided crossing: 
\begin{eqnarray}
\nonumber i\dot{a}&=& \epsilon (t) a+gb , \quad \epsilon (t)=-\beta t+\frac{1}{\kappa} t^2+\ldots\\
i\dot{b}&=& \beta tb + ga,
\label{mod3}
\end{eqnarray}
where $\kappa$ is half of the radius of the curvature (in the energy space) of the diabatic energy near the curve crossing point. It is easy to verify that LZC model corresponds to $\kappa = k/(\beta^{3/2})$. 
Equation (\ref{bp}) shows that if $k \rightarrow \infty$, i.e. if the level crossing is linear, LZC solution transfers into the standard Landau-Zener formula. Solution of LZC shows that when $\kappa$ is finite, the limit of large $g$ and the adiabatic limit do not generally coincide. 

Figure \ref{cross}({\it Left})  shows the effect of the curvature radius of the level. It clearly demonstrates that  the Landau-Zener formula has only a limited range of validity, restricted to large $k$.   The deviations from the Landau-Zener prediction become particularly strong at $k^2<1$. In fact, the simplicity of Eq. (\ref{bp}) and the fact that the transition probability in the LZC-model depends on two rather than one independent parameters suggest that Eq. (\ref{bp}) can be a better approximation at intermediate sizes of the curvature and the level coupling than the Landau-Zener formula.  Figure \ref{check2}({\it Left}) confirms our expectation for the case of quadratically decaying diabatic energy $\epsilon=1/t^2$. Although  Eq. (\ref{bp}) is not much more complex than the Landau-Zener formula, Fig. \ref{check2}({\it Left}) shows that it represents a substantial improvement at intermediate coupling strengths. 
Figure \ref{cross}({\it Right}) also shows results of the numerical check of Eq. (\ref{bm}) at $\beta<0$.

%%%%%%%%%%%%%%%%%%%%%%%%%%%%%%%%%%%%%%%%%%%%%%%%%%%%%%%%%%%%%%%%%%%%%%%%%%%%%%%%%%%%%%%%%%%
\begin{figure}%[!htb]
\scalebox{0.35}[0.35]{\includegraphics{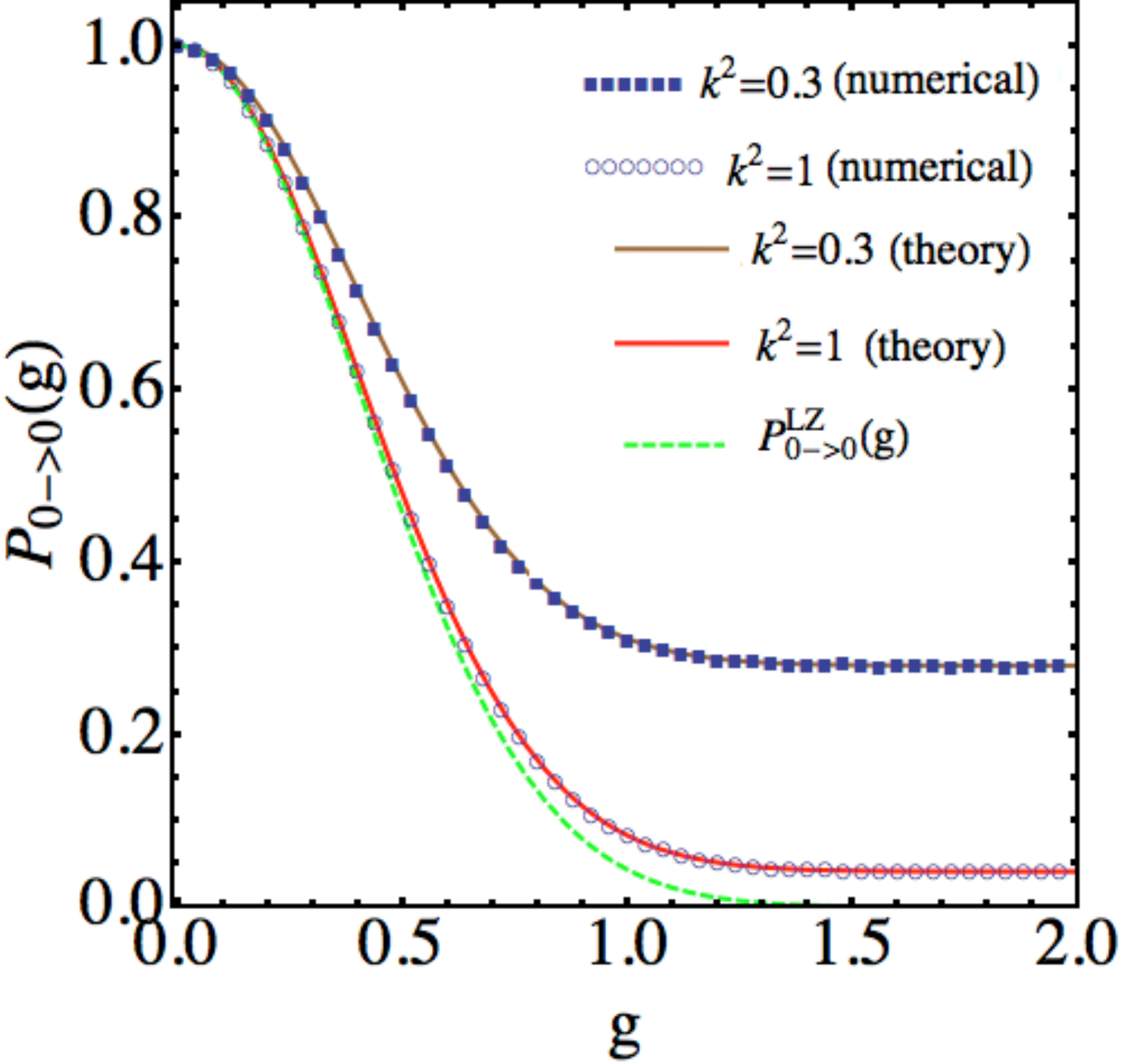}}
\scalebox{0.35}[0.35]{\includegraphics{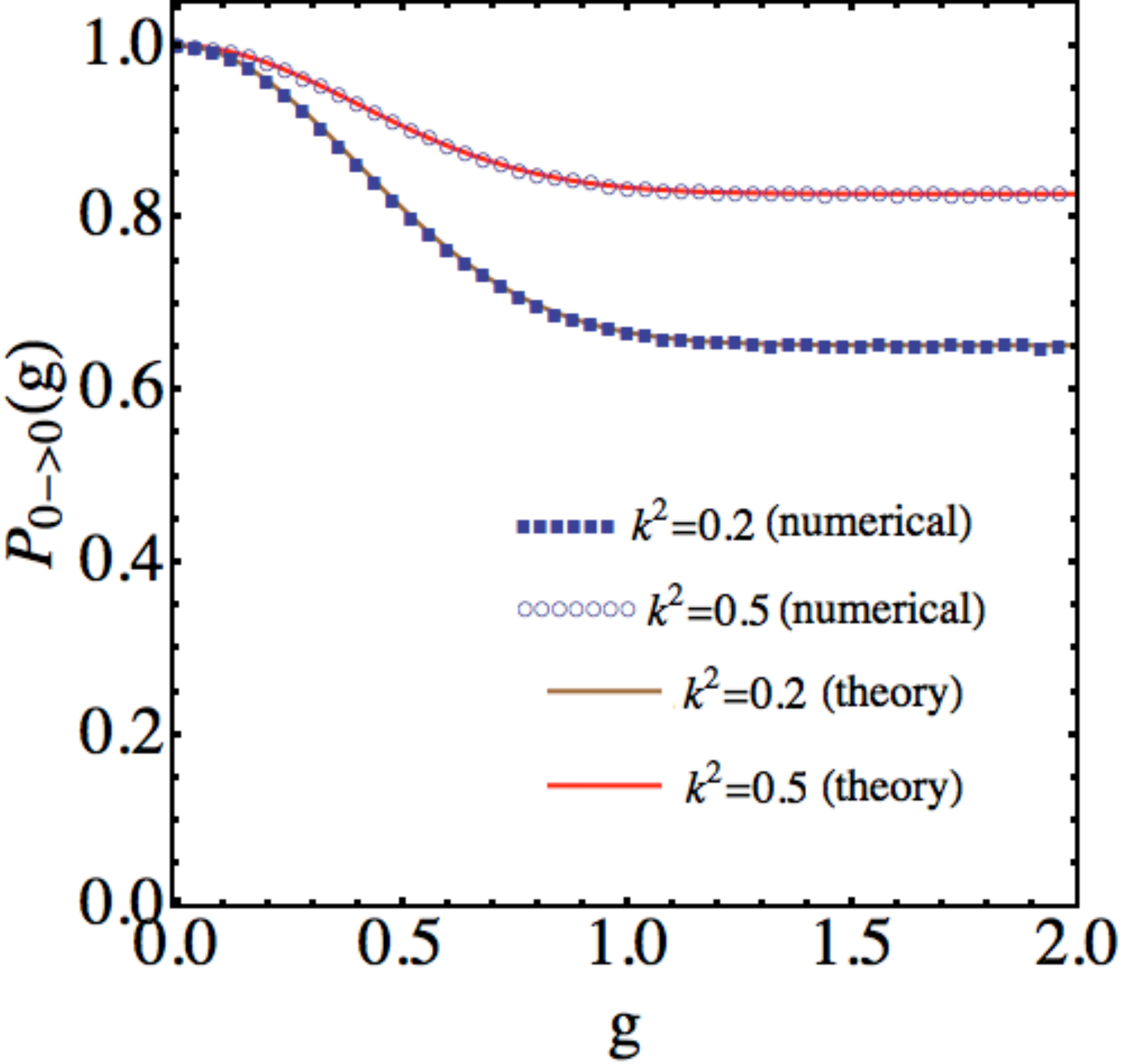}}
\hspace{-2mm}\vspace{-4mm}   
\caption{{\it Left:} Probability to stay at level $0$ at time $t\rightarrow +\infty$  for $\beta =1 >0$. Solid curves are theoretical predictions by Eq. (\ref{bp}). Dashed curve is the Landau-Zener prediction obtained by linearizing time-dependence of diabatic energies near the avoided crossing point. Discrete dots represent the results of the numerical calculations. {\it Right:} Probability to stay at level $0$ at time $t\rightarrow +\infty$  for $\beta =-1 <0$. Solid curves are theoretical predictions by Eq. (\ref{bm}). Discrete dots represent the results of the numerical calculations.
%., as Hermitian Hamiltonians only have real eigenvalues.
} \label{cross}
\end{figure}
%%%%%%%%%%%%%%%%%%%%%%%%%%%%%%%%%%%%%%%%%%%%%%%%%%%%%%%%%%%%%%%%%%%%%%%%%%%%%%%%%%%%%%%%%%%

%%%%%%%%%%%%%%%%%%%%%%%%%%%%%%%%%%%%%%%%%%%%%%%%%%%%%%%%%%%%%%%%%%%%%%%%%%%%%%%%%%%%%%%%%%%
\begin{figure}%[!htb]
\scalebox{0.35}[0.35]{\includegraphics{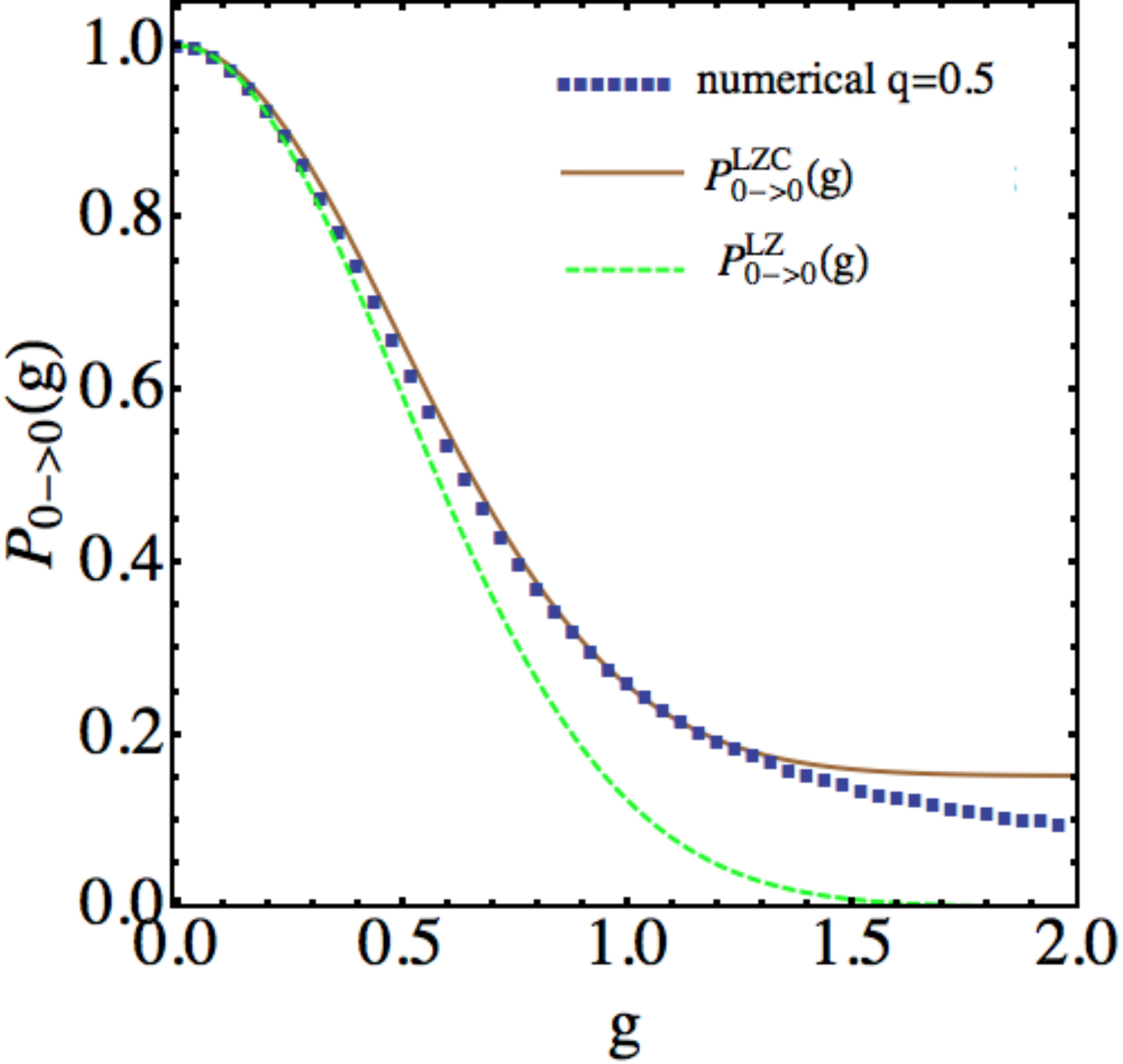}}
\scalebox{0.35}[0.35]{\includegraphics{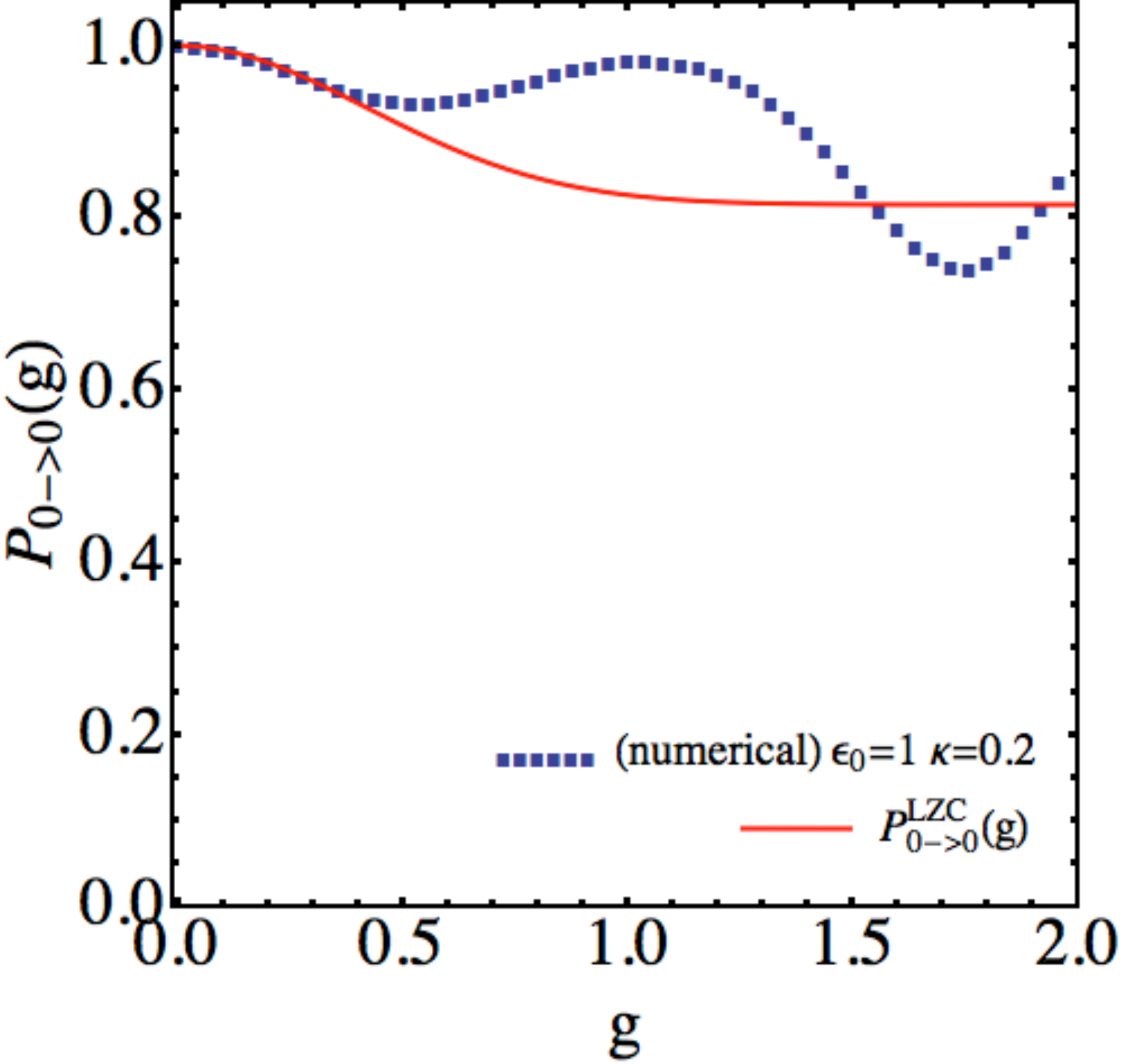}}
\hspace{-2mm}\vspace{-4mm}   
\caption{{\it Left:} Comparison of approximations based on LZ and LZC
  models for a model of an avoided level crossing. Discrete points
  represent results of numerical calculations of the probability to
  stay at the level $0$ at time $t\rightarrow +\infty$  in the model
  of quadratic decay of $0$-th diabatic level: $\epsilon_0(t) =
  q/t^2$, and $\beta=1/2$ in Eq. (\ref{mod3}). Solid and dashed curves
  are the theoretical predictions of, respectively,  Eq. (\ref{bp})
  for LZC and the Landau-Zener formula in which parameters are tuned
  to produce the same level crossing rate and curvature as defined in
  Eq. (\ref{mod3}). {\it Right:} Comparison of approximations based on
  LZC model and numerical results  for the model with quadratic
  diabatic energy of the $0$-th level: $\epsilon (t) =
  \epsilon_0+t^2/(2\kappa)$. For the latter model, discrete points
  represent results of numerical calculation of the probability to
  stay at level $0$ at time $t\rightarrow +\infty$ if the evolution
  starts at $t\rightarrow - \infty$ (not at $t=0$).  Solid line is the
  theoretical prediction of  Eq. (\ref{bm}) for LZC which parameters
  are tuned to produce the same values of $\epsilon_0$ and $\kappa$ in Eq. (\ref{mod4}). 
%., as Hermitian Hamiltonians only have real eigenvalues.
} \label{check2}
\end{figure}
%%%%%%%%%%%%%%%%%%%%%%%%%%%%%%%%%%%%%%%%%%%%%%%%%%%%%%%%%%%%%%%%%%%%%%%%%%%%%%%%%%%%%%%%%%%

\subsection{A model of avoided diabatic level crossing}

If diagonal elements of the Hamiltonian do not intersect but pass close to each other, one can speak about nonadiabatic transitions at avoided diabatic level crossing. 
Two-state models of avoided diabatic level crossings usually can be expressed in the form 
\begin{eqnarray}
\nonumber i\dot{a}&=& \epsilon (t) a+gb , \quad \epsilon (t)=\epsilon_0+ \frac{1}{\kappa} t^2+\ldots, \quad \epsilon_0>0, \\
i\dot{b}&=& ga.
\label{mod4}
\end{eqnarray}
The minimal set of important parameters for such models are the minimal distance between diabatic levels $\epsilon_0$, the relative curvature  $\kappa$ at the moment of reaching this distance  (which characterizes the rate of divergence of levels), and the coupling between the diabatic states $g$. 
In case of LZC, a simple phase transformation  maps this model at $\beta < 0$ to Eq. (\ref{mod4}) with $\epsilon(t)  = |\beta| t + k^2/t $, as illustrated in Fig. \ref{avoid}.

%%%%%%%%%%%%%%%%%%%%%%%%%%%%%%%%%%%%%%%%%%%%%%%%%%%%%%%%%%%%%%%%%%%%%%%%%%%%%%%%%%%%%%%%%%%
\begin{figure}%[!htb]
\scalebox{0.35}[0.35]{\includegraphics{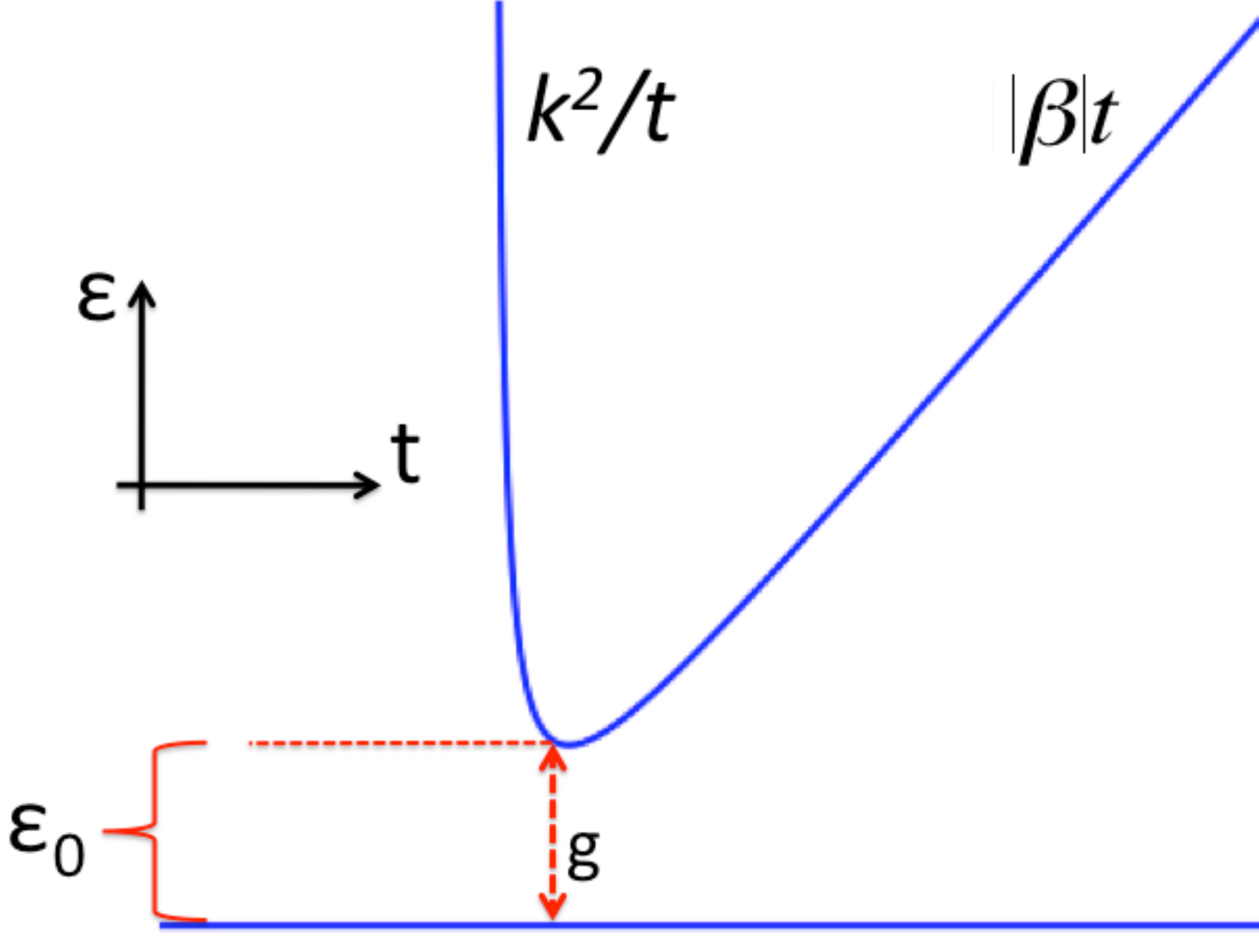}}
\hspace{-2mm}\vspace{-4mm}   
\caption{Avoided diabatic level crossing in a two-state LZC model.
%., as Hermitian Hamiltonians only have real eigenvalues.
} \label{avoid}
\end{figure}
%%%%%%%%%%%%%%%%%%%%%%%%%%%%%%%%%%%%%%%%%%%%%%%%%%%%%%%%%%%%%%%%%%%%%%%%%%%%%%%%%%%%%%%%%%%

%%%%%%%%%%%%%%%%%%%%%%%%%%%%%%%%%%%%%%%%%%%%%%%%%%%%%%%%%%%%%%%%%%%%%%%%%%%%%%%%%%%%%%%%%%%
%\begin{figure}%[!htb]
%\scalebox{0.3}[0.3]{\includegraphics{check2-2.pdf}}
%\hspace{-2mm}\vspace{-4mm}   
%\caption{Probability to stay at level $0$ at time $t\rightarrow +\infty$  for $\beta =-1 <0$. Solid curves are theoretical predictions by Eq. (\ref{res}). Discrete dots represent the results of the numerical calculations.
%., as Hermitian Hamiltonians only have real eigenvalues.
%} \label{check3}
%\end{figure}
%%%%%%%%%%%%%%%%%%%%%%%%%%%%%%%%%%%%%%%%%%%%%%%%%%%%%%%%%%%%%%%%%%%%%%%%%%%%%%%%%%%%%%%%%%%

Parameters $\epsilon_0$, $\kappa$  are straightforward to derive for the LZC model: 
\begin{equation}
\epsilon_0=2k\sqrt{|\beta|}, \quad \kappa=k/(|\beta|^{3/2}).
\label{par}
\end{equation}
Eq. (\ref{par}) can be inverted,
\begin{equation}
|\beta|=\sqrt{\frac{\epsilon_0}{2\kappa}}, \quad k^2=\epsilon_0/(2  \sqrt{|\beta|}) .
\label{par2}
\end{equation}

Unfortunately, it seems that LZC model with parameters  (\ref{par2}) cannot be used as a good approximation to an arbitrary model
of avoided diabatic transitions with the same minimal distance and curvature because of a strongly skewed shape of the curve $\beta t + k^2/t$. Figure \ref{check2}({\it Right})  compares probabilities to remain at the $0$-th level in a model  with a quadratically changing diabatic energy and the LZC model with the same values of basic parameters provided by (\ref{par2}). Reasonably good agreement is found only at relatively small values of $\kappa$ and $g$.

Finally, we note that the theory of nonadiabatic transitions in two-state systems has been very well developed. LZC is only one of many useful solvable systems in this field \cite{rozen,nikitin,osherov}. Moreover, numerical simulations of 
such systems is a trivial exercise. The two-state LZC is
a solvable model with a constant off-diagonal coupling, which is,
sometimes, a required assumption. The simplicity of its transition
probability formulas makes  the LZC model a useful alternative to the
LZ-formula in cases when considerable accuracy is not required but
rather a quick nonperturbative estimate of a transition probability is needed.

%%%%%%%%%%%%%%%%%%%%%%%%%%%%%%%%%%%%%%%%%%%%%%%%%%%%%%%%%%%%%%%%%%%%%%%%%%%%%%%%%%%%%%%%%%%
%\begin{figure}%[!htb]
%\scalebox{0.4}[0.4]{\includegraphics{check3.pdf}}
%\hspace{-2mm}\vspace{-4mm}   
%\caption{Comparison of approximations based on LZC model and numerical results  for the model with quadratic diabatic energy of the $0$-th level: $\epsilon (t) = \epsilon_0+t^2/(2\kappa)$. For  the latter model, discrete points represent results of numerical calculation of the probability to stay at level $0$ at time $t\rightarrow +\infty$ if the evolution starts at $t\rightarrow - \infty$.  Solid line is the theoretical prediction of  Eq. (\ref{res}) for LZC which parameters tuned to produce the same values of $\epsilon_0$ and $\kappa$. 
%., as Hermitian Hamiltonians only have real eigenvalues.
%} \label{check3}
%\end{figure}
%%%%%%%%%%%%%%%%%%%%%%%%%%%%%%%%%%%%%%%%%%%%%%%%%%%%%%%%%%%%%%%%%%%%%%%%%%%%%%%%%%%%%%%%%%%

\end{widetext}

\end{document}